\documentclass{aip-cp}
\usepackage[numbers]{natbib}
\usepackage{rotating}
\usepackage{graphicx}

\begin{document}

\title{Potential for biomedical applications of Positron Annihilation Lifetime Spectroscopy (PALS)}

\author{Ewelina Kubicz\corref{cor1} for the J-PET Collaboration}
\affil[aff1]{Faculty of Physics, Astronomy and Applied Computer Science, Jagiellonian University, S. {\L}ojasiewicza 11, 30-348 Krak{\'o}w, Poland}
\corresp[cor1]{Corresponding author: ewelina.kubicz@doctoral.uj.edu.pl}

\maketitle

\begin{abstract}
	Positron Annihilation Lifetime Spectroscopy (PALS) allows examining structure of materials in nano and sub-nanometer scale. This technique is based on the lifetime and intensity of ortho-positronium atoms in free volumes of given structures. It is mostly used for studies in material sciences, but it can also be used for in vivo imaging of the cell morphology as proposed in \citep{pet}, \citep{patent}. Cancer cells are characterized by an altered macro structure in comparison to normal cells, thus the main objective of these studies is to compare if these differences can be detected on sub-nanometer level and therefore allows to distinguish between normal and cancer cells with application of PALS technique. 

	This perspective will allow for simultaneous determination of early and advanced stages of carcinogenesis, by observing changes in biomechanical parameters between normal and tumour cells, and standard PET examination. Such simultaneous PET imaging and PALS investigations can be performed with the Jagiellonian Positron Emission Tomograph (J-PET) which is a multi-purpose detector used for investigations with positronium atoms in life-sciences as well as for development of medical diagnostics. 
\end{abstract}

\section{INTRODUCTION}
Proposed investigation mainly concerns the studies of the cellular organization and structure at nano and sub-nanometer level and its relation to the cell morphology and signalling. Correlations between cellular processes such as proliferative activity and proapoptotic sensitivity (activity) and Positron Annihilation Lifetime Spectroscopy (PALS) parameters can be determined. Different living normal and cancer cells and tissues will be investigated in order to connect PALS parameters to carcinogenesis and metastatic processes. Cancer cells differ from normal ones by their structure and organelles functionality. These cells are irregular in shape, have a reduced amount of cytoplasm, a bigger and often multiplied nuclei. Organelles such as the Golgi apparatus, mitochondria or endoplasmic reticulum become disorganized or have impaired function. Even though main factors responsible for DNA mutations leading to cancer have been broadly studied over the last few years, there is slight evidence about the molecular structure of the cell, their ultrastructure and the difference between normal cells and cancer cells in proliferation, protein and lipid synthesis and intracellular transport. 

	In our studies we test the hypothesis \citep{pet}, \citep{patent} that cancer cells differ in their sub-nanostructure, molecular interactions or activity, which are specific for cancer. These changed molecular interactions may influence their molecular properties that can be detected as the changed PALS signal. Such detailed research in this area have not been conducted so far, mostly due to the limitations of experimental technique currently used in cell biology.  Recently, the optical super resolution microscopy techniques like Stimulated Emission Depletion (STED) \citep{micro1} or Photoactivated Localization Microscopy (PALM) \citep{micro2} were developed to overcome the barrier of optical resolution and allow to study living biological objects with resolution between $1 - 100~nm$. However, those techniques cannot report on the dynamic behavior of the cellular matter and mostly they require usage of fluorescent labels. When methods like Scanning Electron Microscopy give information of biological specimens at the nano and sub-nanometer scale, but for such accuracy it acquires fixed or dehydrated samples, thus all these studies can be performed only in vitro. Observed changes in biomechanical parameters between normal and tumor cells can be correlated with parameters obtained from PALS technique. Combining this technique with J-PET detector \citep{pet10}, \citep{pet3}, \citep{pet6}, \citep{pet8}, \citep{pet11}, \citep{pet12} would make possible in vivo studies of human cells and tissues in nanometer scale and could be used as additional diagnostic parameter \citep{pet}, \citep{patent}.

\section{DIFFERENCE BETWEEN STRUCTURE AND METABOLISM IN NORMAL AND CANCER CELLS}

Cells and tissues are inherently very complex in their structure. Main research in cell biology and biophysics aims to explain and understand how living processes, which occur on the cellular level are regulated in different spatial and temporal domains. The main challenge in these studies is that the living objects are built of different cellular and molecular structures and chemical compounds, which function depend on many environmental and intracellular factors. Many studies are carried out in the macro and micro spatial distribution but there are only few of them that investigate the regulation of cellular processes in the context of changes at a nano- and sub-nanometer scale of the living objects. The goal of this research is to study normal and cancer cells, as well as compounds building their structure, and cells interactions with extracellular matrix into nano scale using the PALS technique, this will allow for non-optical investigation of the biological structures.

	Studies conducted by the J-PET collaboration aim at bringing the new insight into the carcinogenesis; not only on the genetic and cell physiology levels (mutagenesis, apoptosis, proliferation etc.), but also on the molecular structure in the nano scale, including temporal and spatial domains in the correlation with the free volumes between molecules. Many superimposed factors are responsible for carcinogenesis. Cancer cells show high variations in their genetic and biochemical parameters. Current research demonstrates that mechanical and structural properties of cytoskeleton, cellular environment and even cell nucleus play a major role in carcinogenesis and metastatic processes. Changes in the cellular membranes or cytoskeleton at the macro scale strongly indicate the changes at the micro scale of the membrane, and this could be caused at even smaller level – in nano or sub-nano scale \citep{cell1}. Cancer cells can be divided into two types: benign and malignant. Benign cells have a similarity to their stem cell and are well differentiated, and therefore tumors are progressing slowly as an encysted change. These tumors do not form metastases. Malignant tumors have less differentiated cells and the ability to infiltrate surrounding tissues and cause metastasis \citep{cell2}.

\section{POSITRON ANNIHILATION LIFETIME SPECTROSCOPY}

Positronium is a hydrogen like atom and, at the same time anti-atom. It is build out of electron and its anti-particle positron.  Its bounding energy is equal to $6.8~eV$ and its diameter amounts to about $0.2~nm$, and hence it is sufficiently small to be trapped in the  volumes of lower electron density, so called free volumes between molecules in various living and non-living matter. The smaller are the free voids between molecules and atoms the shorter the o-Ps lifetime. This is due to the pick-off processes in which positron from the positronium may annihilate with one of the surroundings electrons. Shortening its lifetime, when embedded in the matter, can be used for studies of nanomorphology of cells in living organisms \citep{pet}, \citep{patent}. Positronium may be formed in tryplet state as para-positronium (p-Ps), with the average lifetime in a vacuum of $\tau_{p-Ps} = 0.125~ns$, or in singlet state as ortho-positronium (o-Ps) with the average lifetime in a vacuum of $\tau_{o-Ps} = 142~ns$. 

	 Based on theory by Tao \citep{pals1} and Eldrup \citep{pals2} correlation between the lifetime of the ortho-positronium atoms with the free voids size can be determined. These changes of the o-Ps lifetime and intensity of its production are crucial for the biological research. What is important, positronium can also be created in liquids such as water what allows to study living cells. The average lifetime of o-Ps in pure liquid water amounts to about $1.8~ns$ \citep{pals3}.

\subsection{Current PALS application in biology}
	There are few works demonstrating the application of PALS technique to study biological structures. Jean and his group \citep{Jean1} described some applications of positron annihilation techniques in biological systems. They focused on studies of healthy and abnormal skin samples and reported that o-Ps lifetime and therefore free volumes are correlated with the level of skin damage. Both intensity and lifetime of o-Ps were found to be significantly lower in samples with basal cell carcinoma (BCC) and squamous cell carcinoma (SCC) than in normal skin samples. These studies were performed on fixed samples as well as in ambient conditions \citep{Jean2},\citep{Jean3}. 
	
	Another example of advanced PALS application in studies of cell culture is the paper by Axpe et al. \citep{Axpe1}, where they employed well defined colorectal cancer cell lines grown in the 3D matrix. Axpe shown how addition of a growth factor $TGF-\beta$ induces changes at the atomic scale in the size of the free volume voids, due to the biological effects. This studies were carried out in 4 C deg. and showed the possibility of using PALS for live cells research. 

	Some recent PALS studies performed by the J-PET collaboration with simple model micro-organisms unicellular yeasts \textit{Saccharomyces cerevisiae} \citep{Kubicz1}, shows the possibility to observe dynamics of the water sorption by lyophilised (freeze - dry yeast). Lifetime of ortho-positronium was found to change from 2.4 to $2.9~ns$ (longer-lived component) for lyophilized and aqueous yeasts, respectively. Also some temporal changes of the o-Ps lifetime indicating reorganization of yeast in the molecular scale in the presence of liquid water was observed.

	Recently, significant differences in PALS parameters between normal and tumor tissues were also observed by the J-PET group in  samples  of  uterine  leiomyomatis  and  normal  myometrium tissues taken from women-patients after surgery \citep{Bozena1}, \citep{Bozena2}. In case of all samples from six patients it was found that mean o-Ps lifetime are of about $2~ns$, while for normal tissues this value is of about $50~ps$ lower. The above discussed results indicate that in all investigations performed to date there is a difference in PALS parameters determined for healthy and cancerous tissues.

\section{o-Ps lifetime as a new diagnostic parameter with J-PET}

	All studies described in previous section brought important information, showing that PALS technique can be applied in studies of biological structures and human cells and tissues, thus has potential as application in cancer diagnostic. Nevertheless, this studies with PAL spectrometer can only be conducted in vitro. With J-PET scanner \citep{pet3}, \citep{pet6}, \citep{pet1}, \citep{pet2}, \citep{pet4}, \citep{pet5}, \citep{pet7} which is a multi-purpose detector used for investigations with positronium atoms in life-sciences as well as for development of medical diagnostics, such simultaneous PET imaging and PALS measurement is possible, therefore J-PET is capable of imaging properties of positronium produced inside the human body \citep{pet}, \citep{patent}. 
	
	Currently detector is calibrated and first PALS studies of porous material (XAD4 polymer) were performed and results are already in press \citep{pet14}. This allows for use of the J-PET scanner for PALS measurement with human tissue. First such studies were performed with fixed in formaldehyde Cardiac Myxoma \citep{Myxoma1}- benign heart tumor- sample. The data are collected in the triggerless mode by the dedicated electronics \citep{pet15} and the digital data acquisition system \citep{pet2}, \citep{pet16}. The analysis is preformed with the root-based framework \citep{pet17}.
 In Fig.1 (right) obtained lifetime spectra with time resolution of about 270~ps (sigma) was acquired. o-Ps lifetime and intensity was calculated by PALS avalanche program \citep{pet7} with results $\tau_{o-Ps} = 2.03(01)$ and $I_{o-Ps} = 25.7(1)\% $ respectively. With this measurement we have proven that J-PET can be used not only for PET imaging but also for PALS studies of human tissue.
	
	\begin{figure}[h]
 \includegraphics[width=0.25\textwidth]{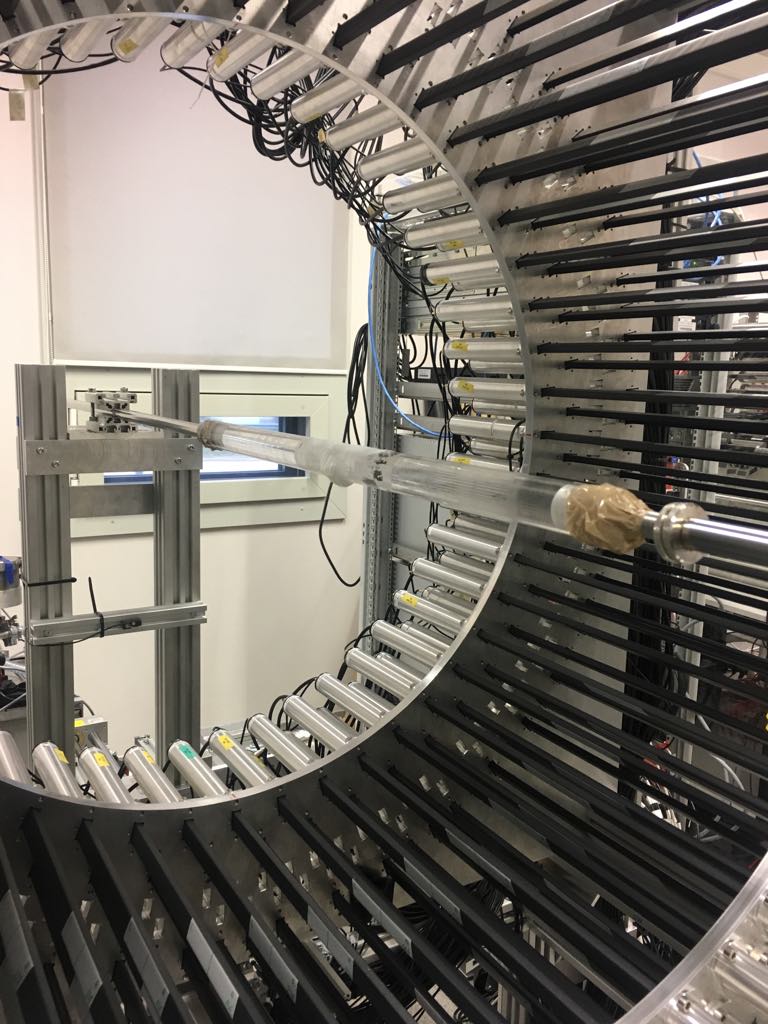}
 \includegraphics[width=0.52\textwidth]{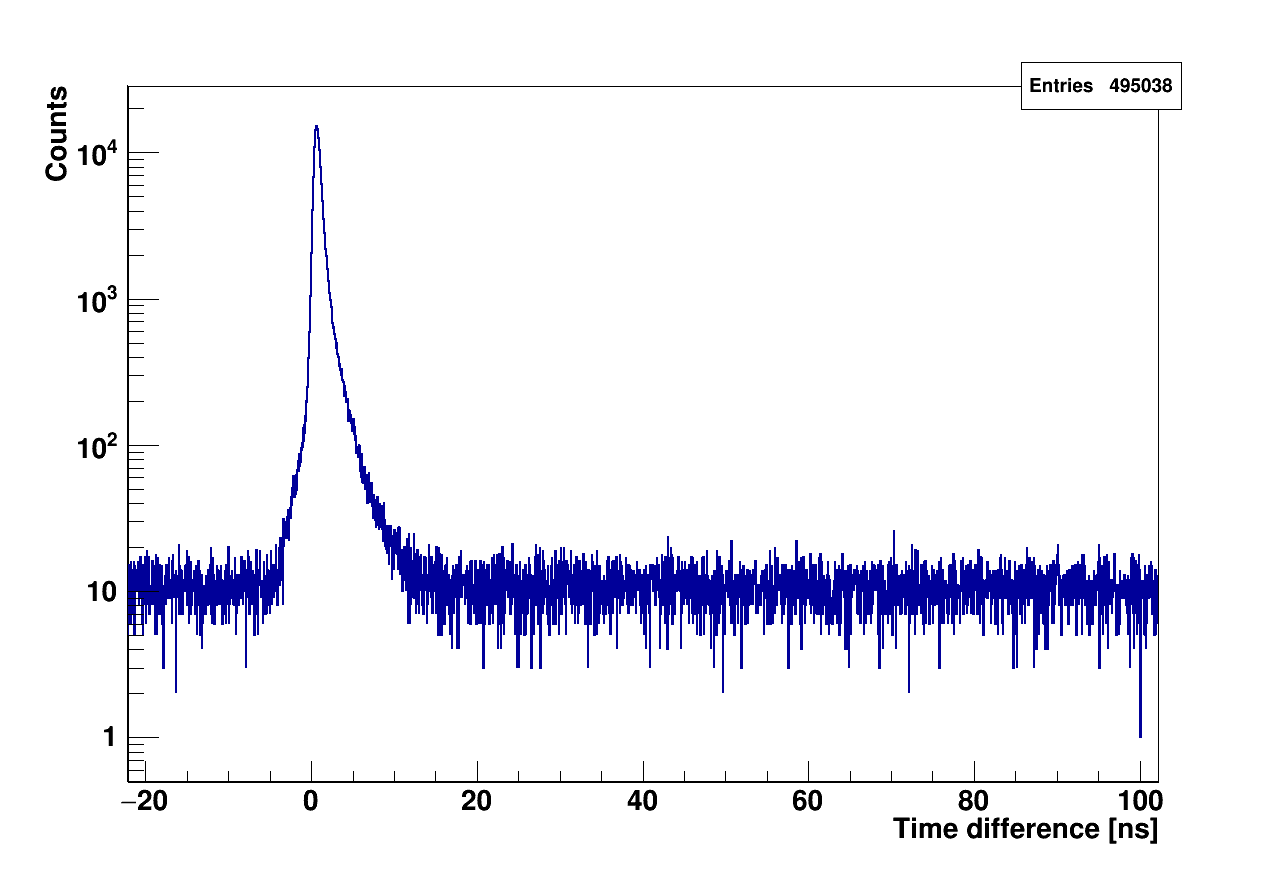}
  \caption{ (left) Photo of J-PET scanner with Cardiac Myxoma inside the holder during measurement. (right) Positronium lifetime spectrum obtained with the J-PET for cardiac myxoma.}
	\label{fig1}
\end{figure}

\section{SUMMARY}
Presented idea of applying PALS for biological studies is quite a new concept, so far few studies, including ones conducted by the J-PET collaboration \citep{Kubicz1}, \citep{Bozena1},\citep{Bozena2}  proven that this technique can be used successfully to differentiate between normal and cancer tissues, therefore can be used in medical diagnostic. Differences between structure and metabolisms of normal and cancer cells, were discussed in this article, but till now there is no explanation what is the reason behind differences observed in PALS. Such information will be crucial for biomedical application. Planned measurements with well defined cell lines possible with use of J-PET scanner will allow to study influence of different factors, like free radicals, reactive oxygen species and glucose metabolism, and thus can lead to some possible model of positronium trapping in living matter and perhaps some new information on carcinogenesis.


%
\section{ACKNOWLEDGMENTS}
The author acknowledges technical and administrative support of A. Heczko, M. Kajetanowicz and W. Migda{\l}. This work was supported by The Polish National Center for Research and Development through grant INNOTECH-K1/IN1/64/159174/NCBR/12, the Foundation for Polish Science through the MPD and TEAM/2017-4/39 programmes, the National Science Centre of Poland through grants no. 2016/21/B/ST2/01222, 2017/25/N/NZ1/00861, the Ministry for Science and Higher Education through grants no. 7150/E-338/M/2018, 6673/IA/SP/2016, 7150/E338/SPUB/2017/1 and 7150/E-338/M/2017, and the Austrian Science Fund FWF-P26783.


\nocite{*}
\bibliographystyle{aipnum-cp}%

\end{document}